# Anomalous compressibility behavior of chromium monoaresenide under high pressure

Zhenhai Yu[1†]*, Wei Wu[2†]*, Jinggeng Zhao[3], Chunyu Li[1], Jinguang Cheng[2],

Lin Wang[1,4,5]*, and Jianlin Luo[2]

[1]Center for High Pressure Science and Technology Advanced Research, Shanghai, 201203, People's Republic of China

[2]Beijing National Lab for Condensed Matter Physics, Institute of Physics, Chinese Academy of Sciences, Beijing 100190, People's Republic of China

[3]Natural Science Research Center, Academy of Fundamental and Interdisciplinary Sciences, Harbin Institute of Technology, Harbin 150080, People's Republic of China

[4]State Key Laboratory of Superhard Materials, Jilin University, Changchun 130012, People's Republic of China

[5]High Pressure Synergetic Consortium, Geophysical Laboratory, Carnegie Institution of Washington, Argonne, Illinois 60439, United States of America

[†]These authors contributed equally to this study and share first authorship.

*Corresponding author, E-mail address: yuzh@hpstar.ac.cn (Zhenhai Yu). welyman@qd-china.com (Wei Wu) and wanglin@hpstar.ac.cn (Lin Wang)

## Abstract

CrAs was firstly observed possessing the bulk superconductivity ($T_c \approx 2$ K) under high pressure (0.8 GPa) in the very recent work (Wei Wu, *et al*. Nature Communications 5, 5508 (2014)). To explore the correlation between the structure and the superconductivity, the high-pressure structure evolution of CrAs was investigated using angle dispersive X-ray diffraction (XRD) method with small steps of ~ 0.1 GPa in a diamond anvil cell (DAC) up to 1.8 GPa. In the pressure range covered by our current experiment, the structure of CrAs keeps stable. However, the lattice parameters exhibit anomalous compression behaviors. With the pressure increasing, the lattice parameters *a* and *c* both show a process of first increasing and then decreasing, and the lattice parameter *b* goes through a quick contract at 0.35 GPa, which suggests a pressure-induced isostructural phase transition occurs in CrAs around this pressure point. Above the phase transition pressure, the axial compressibilities of CrAs present remarkable anisotropy. The compressibilities along the *a*- and *c*-axis are about an order of magnitude smaller than that along the *b*-axis, which is closely related to the different stacking modes in $CrAs_6$ octahedron along different crystallographic axes. A schematic band model was used for addressing above anomalous compression behavior in CrAs.

**1. Introduction**

The discovery of the superconductivity ($T_c$ = 26 K) in ZrCuSiAs-type LaFeAs($O_{1-x}F_x$) [1] had inspired widely experimental and theoretical researches on the quaternary "1111" compounds $R$FeAsO, where *R* represents a lanthanide (La, Ce, Nd, etc.) [2-5]. From the point of view of crystallography, the new discovered iron-based superconductor exhibits quasi-two-dimension structural character at ambient conditions. Like the $CuO_2$ plane in the copper oxide high-temperature superconductors, the $Fe_2As_2$ layers are the conduction planes for charge carriers, the other building blocks are the charge reservoir layers, dominating the carrier density or the chemical potential. For achieving higher superconducting transition temperature, the fundamental structure factors are of high interest for scientific researches. The preceding investigation suggests that the symmetry of the tetrahedral $FeAs_4$[6], the FeAs interlayer spacing[7] and the arsenic atomic height from the Fe-plane[8,9] are all crucial. Just as CuO which plays as a structural proxy for the layered cuprate perovskite family of compounds, the binary transition metal monoarsenides (TAs) could also be considered as the same role. It sheds a light on the exploring of novel superconductors such as the recently discovered CrAs superconductor[10,11].

The transition metal monoarsenides CrAs and MnAs were highly studied during the past decades due to their complex magnetism and magnetcocaloric effects[12,13]. For FeAs, the antiferromagnetic order was observed below 70 K[14], and the renewed interest in FeAs is due to the discovery of the iron-arsenide based superconductors. The CrAs crystallizes into a MnP-type structure as shown in Figure 1 (a) under ambient conditions. The neutron diffraction studies for CrAs showed that it is an double helical type antiferromagnet with the Néel temperature ($T_N$) of ~ 260 K[15]. The magnetic transition of CrAs at $T_N$ occurs as the first order transition companying an abrupt variation of the lattice parameters. And the lattice parameters $a$ and $c$ (space group with *Pnma* setting) decrease while the lattice parameter *b* increases at $T_N$ as the temperature decreasing[16]. In our previous study[17], the high quality single crystal CrAs was grown using the Sn flux method. And a clear magnetic transition was observed with a sharp drop in the resistivity and the susceptibility at about 270 K. The extensive investigations on the CrAs were also focused on exploring the magnetism changing by engineering the structure of CrAs. These approaches are doping the phosphorus elements such as P or Sb into the As site[18-20], or doping the transition metals such as Ti, V or Fe into the Cr site[21-23]. As a perfect source to inject spin polarized current, the half-metallic CrAs thin film ferromagnets [24-26] compatible with the semiconductor materials has also attracted increasing interest for spin electronics[27-29].

Besides the chemical doping, the temperature and pressure as the conventional thermodynamics parameters can also tailor the electronic and the magnetism properties by changing the interatomic distance and consequently resulting in the variation of the lattice parameters. The lattice parameters of $CrAs_{1-x}P_x$ as a function of the compositions have been investigated by Kanaya *et al.*[15] using the XRD technique. The experimental results indicate that the lattice parameters *a* and *c* are almost constant, while the *b* decreases sharply to $x = 0.05$. The mechanism of this anomalous variation is still not fully understood. All the three lattice parameters *a*, *b* and *c* decease almost linearly with *x* in the range of $x > 0.2$, which could be reasonably attributed to the smaller ironic radii of the P element compared to the As element. CrAs, which possess the MnP-type structure at room temperature, transforms to the NiAs-type structure at high temperature of 1173(20) K[30-32]. Pressure dependence of $T_N$ for CrAs has been investigated by Keller *et al.*, and showed that the $T_N$ decreases as the pressure increasing[33,34]. The measurement for the electric conductivity of CrAs at 4.2-350 K under high pressure up to 0.9 GPa was performed by Zavadskii *et al.*[35]. It was shown that CrAs has the metallic conductivity in the entire range of pressures and temperatures. Recently, Shen *et al.* investigated the structural and magnetic phase diagram of CrAs as a function of the temperature and pressure[36]. The experimental results indicated that a spin reorientation was induced above the critical pressure (~0.6 GPa) and the nearest-neighbor bond lengths of the Cr atoms were significantly reduced.

In the past few decades, most of the reported literatures were focused on investigating the chemical doping, temperature or pressure effect on the structural phase transitions or on the magnetic and electric properties of CrAs. However, the mechanism that the lattice parameters of CrAs show a juncture at $T_N$ is still not fully understood and few documents do provided the information on its structural evolution under high pressure, which is essential to understand the relation between the structure and the properties of CrAs under compression. To gain more information on the structure of CrAs, in this paper, the in situ high pressure XRD experiment on CrAs has be undertaken using an angle dispersive XRD technique with very small pressure steps. It was found the lattice parameters present anomalous compression behaviors under high pressure according to the Rietveld refinement results.

## 2. Experiment

The high pressure synchrotron XRD experiment was carried out using a symmetric diamond anvil cell with 400 μm culet. The T301 stainless steel with a thickness of 310 μm and a pre-indentation thickness of 50 μm was served as the gasket. The 150 μm diameter sample chamber was filled with a mixture of the CrAs powder, a ruby chip, and some silicone oil as the pressure transmitting medium. Angle dispersive powder XRD patterns were taken with a Mar2048 detector using synchrotron radiation beams monochromatized to a wavelength of 0.4112 Å at X17C beamline of the National Synchrotron Light Source(NSLS), Brookhaven National Laboratory (BNL). An independent angle dispersive XRD study on CrAs under high pressure was also carried out at beamline BL15U1 of Shanghai Synchrotron Radiation Facility (SSRF) using the wavelength of 0.6199 Å and CCD detector. The two-dimension image plate patterns were converted to the one-dimension intensity versus degree data using the Fit2D software package[37]. The experimental pressures were determined by the pressure-induced fluorescence shift of ruby[38]. The XRD patterns of CrAs were analyzed with Rietveld refinement using the GSAS program package[39] with a user interface EXPGUI[40].

## 3. Experimental results

### 3.1 The crystal structure of CrAs at various temperatures and ambient pressure

As mentioned, at ambient pressure, CrAs possesses two crystallographic polymorphisms as shown in Fig. 1. One is the MnP-type orthorhombic structure (Fig.

1 (a)) stable at ambient conditions with space group *Pnma*, which has four formula units per unit cell. The Cr and the As ions locate at the 4c Wyckoff positions (0.004, 0.25, 0.211) and (0.199, 0.25, 0.584), respectively. It could be found the coordinates of the Cr and the As ions follow the empirical constraint as $x_T \approx 0$, $z_T \approx x_X \approx 0.20$ and $z_X \approx 7/12$ (T = Transition metal, X = Phosphorus element), which are obeyed for most MnP-type compounds. The other high temperature crystallographic polymorphism is the NiAs-type hexagonal structure (Fig. 1 (b)) with space group $P6_3/mmc$, two formulas contained per unit cell. The Cr and the As ions locate at the 2a and the 2c Wyckoff positions, respectively.

**3.2 High pressure angle dispersive XRD patterns of CrAs**

Fig. 2 (a) demonstrates the selected patterns of the angle dispersive XRD results of CrAs under various pressures. The sample was pressurized in small steps of ~ 0.1 GPa. At the first glance of Fig. 2(a), it can be clearly observed the Bragg peaks shift towards higher angles owing to the lattice compression without any modification of the overall peak patterns and no new diffraction peaks were observed in the XRD patterns up to 1.81 GPa, except for apparent separation of the diffraction peaks. The crystal structure of CrAs remains the same space group in the present investigated pressure range. To get a further insight into the pressure effect on the XRD patterns of CrAs, the local diffraction peaks are magnified in the plot of Fig. 2 (b). It can be seen that the diffraction peak (102) related to the lattice parameters *a* and *c* shifted to the lower diffraction angle (or larger *d*-spacing) as the pressure increasing from 0 to 0.57

GPa. This phenomenon is opposite to the conventional compression law. As the pressure further increasing, the diffraction peak (102) exhibits normal compression behavior and shifts to the higher diffraction angle. The diffraction peak (111) corresponding to the variations of all the lattice parameters *a, b* and *c* displayed more pressure dependence.

Fig. 3 shows the finial refinement of the diffraction pattern at ambient conditions collected at X17C beamline of NSLS at BNL using the *Pnma* model referenced from Ref. 14. The tick marks below the diffraction pattern indicate the location of the calculated diffraction peaks in the plot. The residual difference between the calculated and the experimental patterns is shown at the bottom of the plot.

GSAS refinements of the XRD patterns for CrAs at different pressures permit us to obtain the pressure dependences of the lattice parameters, which are shown in Fig. 4. In general, pressure shortens the interatomic distance of the material and the lattice parameters contract accordingly. However, it could been found from Fig. 4 (a) and (c) that the lattice parameters *a* and *c* increase with pressure increasing from 0 to 0.57 GPa. It is thermodynamically impossible for a closed system to have a negative volumetric compressibility[41]. The quick shortening of the lattice parameter *b* results in the lattice volume decreasing under external applied pressure. And the lattice parameter *b* has a steep change at 0.35 GPa. We suggested that this phenomenon is a pressure-induced isostructural phase transition. The changing trend of the unit cell volume is consistent with that of the lattice parameter *b*. It should be emphasized no space group change occurs for this isostructural phase transition, except the anomalous variation of the lattice parameters under high pressure. A detailed discussion of this phase transition will be given below. To testify the credibility of this observed phenomenon, another independent synchrotron angle dispersive XRD experiment was performed and the experimental results exhibit the same phenomenon. The detailed information of the angle dispersive XRD experiment of CrAs is shown in the supplementary file.

**4. Discussion.**

The axial compressibility of CrAs is remarkably anisotropic after the isostructural phase transition point, and the relative axial compressibilities along the $a$-, $b$- and $c$-axis are 0.14%, 2.85% and 0.097%, respectively. The compressibilities along the $a$- and $c$-axis are more than an order of magnitude smaller than that along the $b$-axis.

CrAs could be considered as a three-dimensional network comprised with $CrAs_6$ octahedra contrast to the quasi-layers in the iron-based superconductor. With regard to the anisotropic axial compressibility of CrAs, it could be explained by the different stacking of the $CrAs_6$ octahedron along the different crystallographic axes. The structure of the Cr centered $CrAs_6$ octahedron viewed from different directions is depicted in Fig. 5. The $CrAs_6$ octahedra are linked by face-connecting along the $a$-axis and edge-shared along the $b$-axis. It is interesting to note that the neighboring $CrAs_6$ layers along the $c$-axis are by half edge-shared and half vertex-shared connections. The average number of the CrAs octahedral layers along the $a$-, $b$- and $c$-axis per unit cell is 2, 1 and 2, respectively. And the thickness of every $CrAs_6$

octahedral layer along the *a*-, *b*- and *c*-axis is 2.8206 (5.6411/2), 3.4797 (3.4797/1) and 3.0980 Å (6.1959/2) Å, respectively. Thus the CrAs$_6$ octahedra layers along *b*-axis are sparser than that of the *a*- and *c*-axis.

The calculated bonding lengths and bonding angles for the environment surrounding the Cr atoms indicate that the CrAs$_6$ octahedron is not a perfect one at ambient conditions. The lengths for different Cr-As bonds are very large as shown in Fig. 6. The more distortion of the CrAs$_6$ octahedron and the shortening of the Cr-As bonds make the CrAs$_6$ octahedron greatly squeezed along the *b* axis under high pressure, which also give rise to the increase of the crystal-field splitting energy. So the lattice parameters *a* and *c* show subtle variation, while the lattice parameter *b* shows great shrink at 0.35 GPa.

The band model of TAs was first proposed by Goodenough *et al.*[42-44], and then developed by Boller and Kallel [45]. Here, we interpret the anomalous compression behaviors of CrAs at~ 0.57 GPa under the frame of the above mentioned band model. The electronic configuration of the Cr and the As atoms is $3d^54s^1$ and $4s^24p^3$, respectively. In the light of the crystal ligand field theory, CrAs crystallizes into the MnP-type structure and every Cr atom has an octahedral coordination with the first neighbors being of As atoms. The *d* states of the Cr atom split into $t_{2g}$ ($d_{xy}$, $d_{yz}$ and $d_{xz}$) and $e_g$ ($d_{z^2}$ and $d_{x^2-y^2}$) manifolds. The $t_{2g}$ states of Cr hybridize with the *p* state of As atoms, forming the bonding and the anti-bonding states. The $e_g$ states of Cr are nonbonding and form the narrow states. At ambient pressure conditions, the energy bands related to the *b*-axis forms an empty band above the Fermi level, in which no electrons filled, no electron repulsion along the *b*-axis. Therefore, the *b*-axis is much more compressible at the onset of pressure applying compared to the *a*- and *c*-axis. As pressure increasing, the width of the whole energy bands in CrAs, including the b-axis related energy bands, is broadened. When the applied pressure is increased to the critical point (0.35 GPa, deduced from the pressure dependence of the lattice parameter (Fig. 4)), the b-axis related energy band just crossed into the Fermi level with partially filled electrons. The electronic repulsion caused the b-axis more incompressible. So the energy band model can successfully addresses the curve for the pressure dependences of the lattice parameter b as shown in Fig.4 (b).

As seen from Fig. 4 that the lattice parameter a and c of CrAs exhibited elongation while lattice parameter b showing quick contraction among 0 ~ 0.57 GPa. The remarkable variation of lattice parameters resulted in reorientation of Cr orbital under high pressure. Consequently, the magnetism of CrAs exhibited changes such as abrupt drop of the magnetic propagation vector at a critical pressure (~ 0.6 GPa)[35].

## 5. Conclusions

We present crystallographic structure studies of chromium monoaresenide at room temperature in a diamond anvil cell with small pressure step using synchrotron X-ray powder diffraction. The lattice parameter of CrAs exhibited anomalous variation as a function of pressure. We suggested this was pressure-induced isostructural phase transition. In the present investigated pressure scope CrAs was remarkably anisotropic. The structure evolution behavior of CrAs under high pressure may shed light on the pressure-induced superconductivity. It's worth noting that the transition pressure (~0.57 GPa) obtained from this work coincides with the magnetic phase transition and the emergence of bulk superconductivity.

**Acknowledgements**

Portions of this work were performed at the BL15U1 beamline, shanghai synchrotron radiation facility in China. The authors would like to thank Shanghai Synchrotron Radiation Source for use of the synchrotron radiation facilities. We acknowledge the National Synchrotron Light Source (NSLS) of Brookhaven National Laboratory (BNL) for the provision of synchrotron radiation facilities beam line X17C. We are thankful for support from COMPRES (the Consortium for Materials Properties Research in Earth Sciences).

**Figure captions**

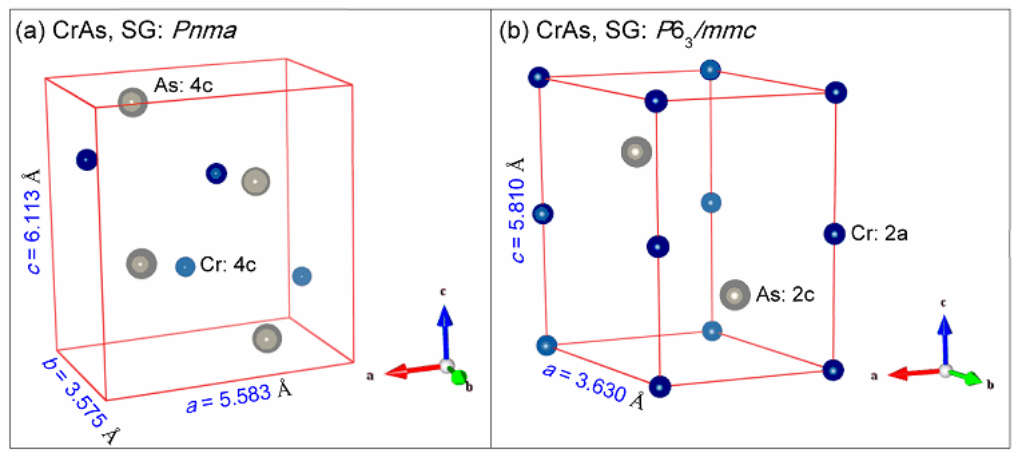

Fig. 1. The crystallographic polymorphisms of CrAs at various temperatures and ambient pressure, (a) the orthorhombic structure with space group *Pnma*, (b) the hexagonal structure with space group *P*6$_3$/*mmc*

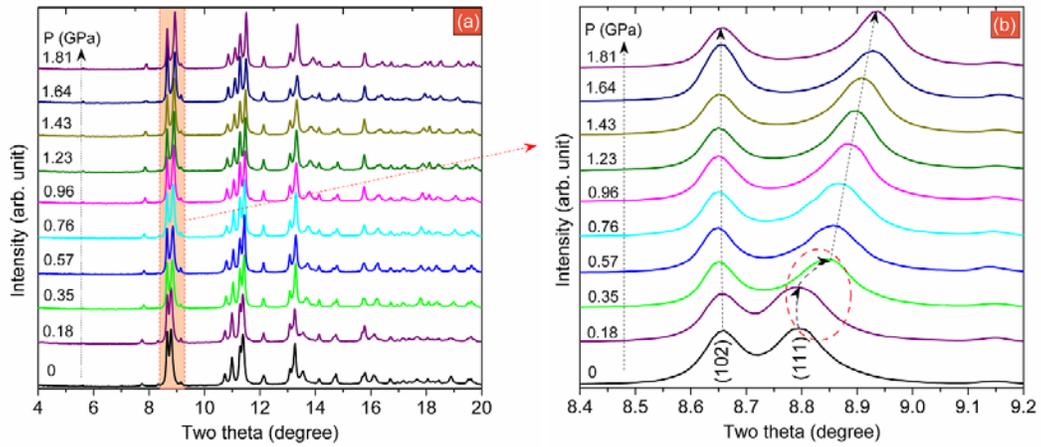

Fig. 2. (a)The selected patterns of the angle dispersive XRD results of CrAs under various pressures. Pressures are indicated in the figure. (b) The local diffraction peaks magnified to highlight its pressure dependence. The dotted lines are guide to the eyes.

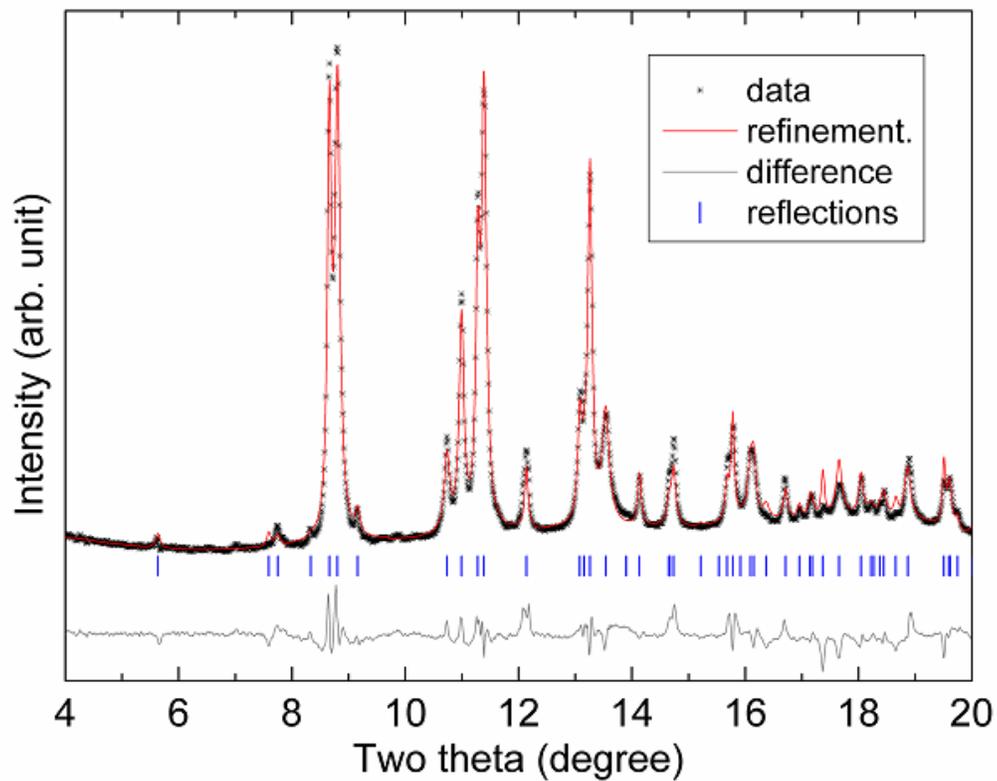

Fig. 3. Refinement of angle dispersive XRD data of CrAs at ambient conditions. Measured (dots) and calculated (solid lines) patterns are shown together with the difference curve (gray lines) and calculated positions of Bragg reflections (tick marks).

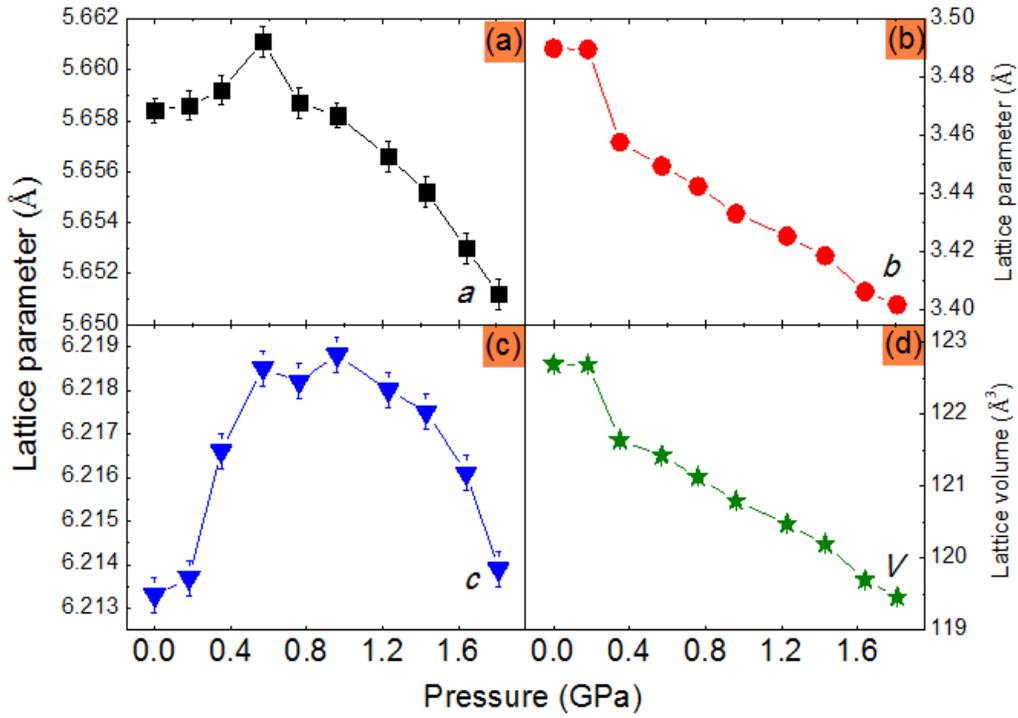

Fig. 4, (a)-(c) Pressure dependences of the lattice parameters *a*, *b* and *c* of CrAs. (d) Lattice volume as a function of pressure for CrAs.

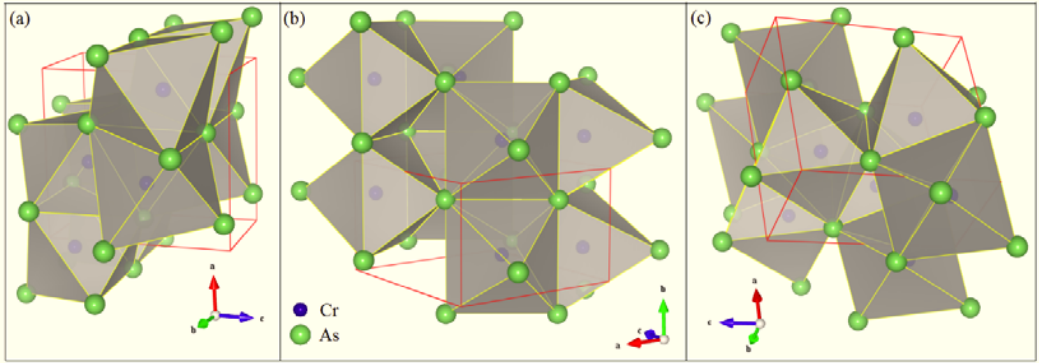

Fig. 5 The schematic of CrAs$_6$ octahedra along *a*-, *b*- and *c* axis of CrAs.

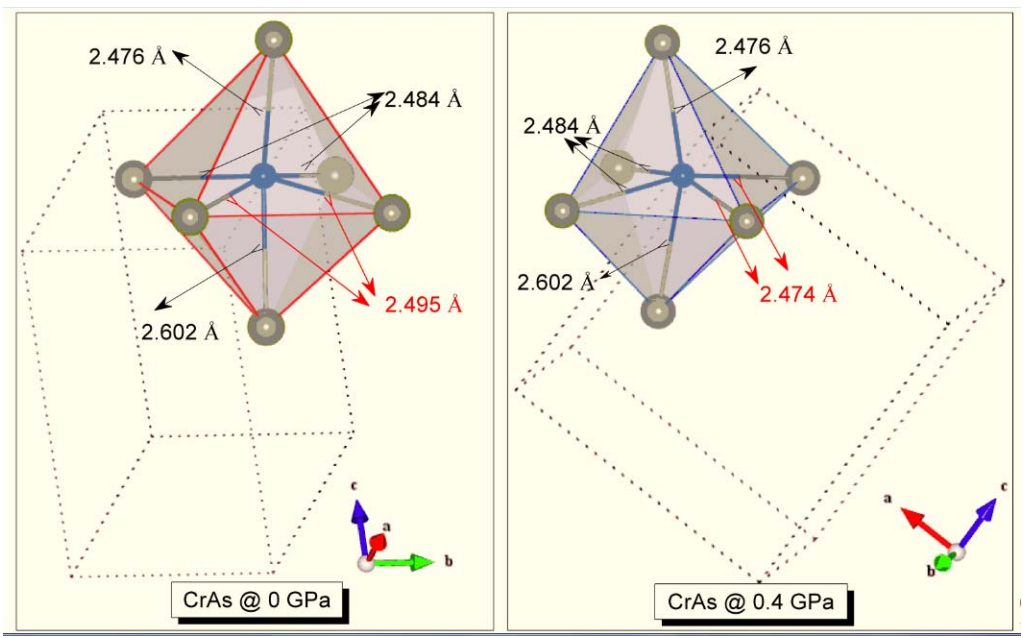

Fig. 6. The bonding lengths of the CrAs$_6$ octahedron under ambient conditions (a) and 0.35 GPa (b), respectively.